# A Trust Domains Taxonomy for Securely Sharing Information: A Preliminary Investigation

N.A.G. Arachchilage and A. Martin

Department of Computer Science
University of Oxford
Wolfson Building, Parks Road
Oxford, UK, OX1 3QD
e-mail: {nalin.asanka; andrew.martin}@cs.ox.ac.uk

## Abstract

Information sharing has become a vital part in our day-to-day life due to the pervasiveness of Internet technology. In any given collaboration, information needs to flow from one participant to another. While participants may be interested in sharing information with one another, it is often necessary for them to establish the impact of sharing certain kinds of information. This is because certain information could have detrimental effects when it ends up in wrong hands. For this reason, any would-be participant in a given collaboration may need to establish the guarantees that the collaboration provides, in terms of protecting sensitive information, before joining the collaboration as well as evaluating the impact of sharing a given piece of information with a given set of entities. In order to address this issue, earlier work introduced a trust domains taxonomy that aims at managing trust-related issues in information sharing. This paper attempts to empirically investigate the proposed taxonomy through a possible scenario (e.g. the ConfiChair system). The study results determined that *Role, Policy*, *Action, Control, Evidence* and *Asset* elements should be incorporated into the taxonomy for securely sharing information among others. Additionally, the study results showed that the ConfiChair, a novel cloud-based conference management system, offers strong privacy and confidentiality guarantees.

## 1. Introduction

Human trust is subjective (Bizer and Oldakowski, 2004, Wang and Emurian, 2005). Therefore, it is worth evalating the human aspect of trust in a system in order to better design trustworthiness systems for eveyone. In our day to day life, we use a wide range of trust decisions. These decisions depend on the specific situation, our experience, subjective preferences, and also the trust relevant information displayed by the system. For example, we might trust to buy a particular item only from sellers on Amazon who have more than 100 positive ratings. Therefore, it is important to display measureable trust characteristics from the system to attract customers for the business.

Information sharing plays an important role in any business process. Organizations and individuals exchange information daily for the purpose of service delivery, communication and collaboration. For example, two law enforcement agencies working on similar cases may require sharing information about the evidence of a crime. However, each agency may need to share such information with only a selected list of other agencies or individuals from these agencies. While participants may be interested in sharing information with one another, it is often necessary for them to establish the impact of sharing certain kinds of information. This is because certain information could have detrimental effects when it ends up in wrong hands. For this reason, any would-be participant in a collaboration may need to establish the guarantees that the collaboration provides, in terms of protecting sensitive information, before joining the collaboration as well as evaluating the impact of sharing a given piece of information with a given set of entities.

Arachchilage, et al. (Arachchilage *et al.* 2013), proposed a concept of a trust domains that aims at managing trust-related issues in information sharing. It is essential for enabling efficient collaborations. Authors introduced a taxonomy for trust domains with measurable trust characteristics, which provides security-enhanced, distributed containers for the next generation of composite electronic services for supporting collaboration and data exchange within and across multiple organisations. Then the proposed taxonomy applied to a possible real world scenario, in which the concept of trust domains could be useful. Kirkham, et al. (Kirkham *et al.* 2013), describes individuals are transient on the Internet, but the data is permanent. Individuals exist only at the outside of the architecture, and behind a browser, application, service, or device. Authors argued that the indispensability of a user-centric architecture where individuals need some kind of unified, permanent, and controllable representation of themselves (Kirkham *et al.* 2013). Therefore, it is important to understand how users trust perceptions towards the taxonomy since it developed for them to securely share the information among the others. On the other hand, they are ultimately responsible of sharing information among others. It can therefore be argued that it's worth understanding their trust perceptions before actually implementing the taxonomy in the real world. Therefore, our purpose in the current study is to validate the taxonomy by exploring users' trust perceptions of the ConfiChair system scenario.

Kelton, et al. (Kelton *et al*. 2008), assert that there is a strong need for empirical research on the end user's trust perception of using systems within the field of information science. So far, there has been little research work on the human aspect of trustworthiness systems designing and modeling (Baptista *et al.* 2008, Donaldson and Fear, 2011, Missier *et al.* 2008 and Weng *et al.* 2007). We know to our cost that none of the existing models or systems has been empirically tested with end-users before their implementation. Furthermore, It has been shown that previous end-user studies underscore the importance of human-centered systems modeling in determining the trustworthiness of systems (Baptista *et al.* 2008, Chapman *et al.* 2010, Gil and Artz, 2007, Hartig and Zhao, 2009, Lauriault *et al.* 2008, Missier *et al.* 2008 and Sexton *et al.* 2004). It is worth empirically testing those systems and models before the implementation takes place since they were designed for end-users. However, there has been little empirical research on exploring those assumptions. The research work reported in this paper attempts to empirically investigate the proposed taxonomy (Arachchilage *et al.* 2013), for trust domains through users based on the ConfiChair system scenario. The concept of trust domains is used for proving a foundation (evidence) for securely sharing information (how, when and with whom) among a group of entities. This enables the parties involved and the observers to understand the level of trust before going ahead with sharing data. Therefore, the current study empirically investigates what key elements that should be incorporated into the trust domains taxonomy for securely sharing information among others. Furthermore, it interprets why these elements should be incorporated into the trust domains taxonomy.

The developed taxonomy enables individuals and organizations to securely collaborate across functions, geographies and corporate boundaries by providing collaborating parties (or participants) the means to create online environments designed to prevent information from leaking and where their resources can be shared as they specify.

The reminder of this paper is structured in the following manner. Section II describes the proposed taxonomy for trust domains. In section III, the proposed taxonomy is applied to a possible scenario (in this case the ConfiChair system scenario). We then discuss the methodology employed in this research to empirically evaluate the taxonomy through the ConfiChair system scenario in section IV. Section V describes the results analysis. Then the trust domains taxonomy is formed through an empirical investigation in section VI. In section VII, a detailed discussion of the findings is presented. Section VIII provides conclusions and opens up opportunities for future work that may extend the research work reported in this paper.

## 2. Proposed trust domains taxonomy

In this section of the paper, we discuss how the models were combined to create the trust domains taxonomy (Arachchilage *et al.* 2013). We illustrate the concepts that can be used to integrate the models and discuss how the semantic gap in the usage of these concepts can be bridged.

### 2.1. Fundamental Concepts and Relations

The proposed model consists of a number of concepts, such that each concept captures a class of things that may either exists in a trust domain, be used to build a trust domain and used within a trust domain. Though all these concepts may be used in different instances of a trust domain, a few of them can be identified as being fundamental to the existence of a trust domain. We identify the *Data, Policy, Controls, Roles, Actions* and *Evidence* as being fundamental concepts necessary to build a trust domain.

As depicted in Figure 1, a *Role* owns *Data* that will exist within a trust domain and establishes a *Policy* that constrains *Actions*.

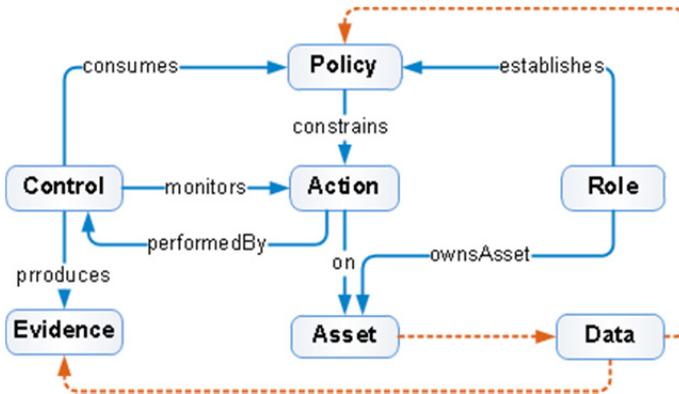

**Figure 1: Fundamental concepts in the trust domain taxonomy**

As mentioned in the fundamental model, a *Role* establishes one or more *Policies* within the domain. However, any given policy can only be established by one *Role*. This means that if two roles establish identical policies, then both policies are treated as a unique entity, which can be linked through the equivalence property.

*Actions* are performed by a given role or by an agent that represents a particular role. These actions are monitored by *Controls* to ensure that the policy is being upheld. These controls produce *Evidence* to indicate that actions have been performed in accordance to the policies. Both *Evidence* and *Policy* can be considered to be a form of data, which can be manipulated in the same way as other data and may be subject to the same information flow restrictions.

## 3. Application Scenario

We consider a scenario, which focuses on the particular cloud computing application of conference management. Existing conference management systems like EasyChair and Editor's Assistant (EDAS) pose the specific security and privacy risks. The Cloud service provider (e.g. the cloud system administrator who administrates the system for all conferences) has access to all the data, and could accidentally or deliberately discloses it to the public. On the other hand, an individual conference chair (who is concerned with a single conference) has access to the data only for the particular conference of which one is chairing. Furthermore, an author or reviewer that chooses to participate in the conference can be assumed to be willing to trust the chair (it can therefore be argued if he didn't, he would not participate); but there is no reason to assume that he trusts or even knows the conference management system provider. Therefore, those conference management systems raise privacy and confidentiality issues. Arapinis, et al. (Arapinis *et al.* 2012), proposed a conference management system called ConfiChair (as shown in Figure 2) to address those issues.

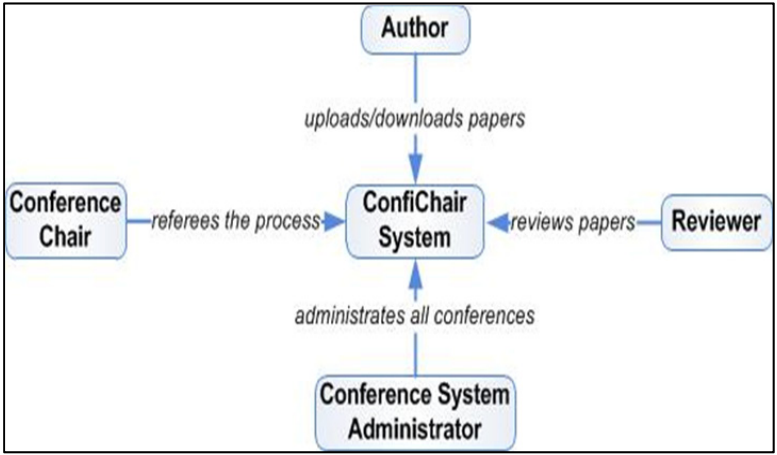

**Figure 2: ConfiChair conference management system**

ConfiChair is a cloud-based conference management protocol, which is proposed to address a set of privacy and confidentiality requirements for conference management. In ConfiChair, authors, reviewers, and the conference chair interact through their browsers with the cloud, to perform the usual tasks of uploading and downloading papers and reviews. Authors claimed that the ConfiChair system offers strong privacy and confidentiality gurantees. Therefore, we use the ConfiChair system as the application scenario to empirically investigate the proposed taxonomy in Figure 1 (Arachchilage *et al.* 2013).

### 3.1. Data Flow Relations & Trust Domains Application

In this section of the paper, we attempt to apply the proposed trust domains taxonomy to the above-mentioned scenario (in this case ConfiChair system scenario). This is achieved by identifying the boundaries that exist with respect to data flows within the setup.

According to the ConfiChair system scenario shown in Figure 2, authors, reviewers, and the conference chair interact through their browsers with the cloud, to perform the usual tasks of uploading and downloading papers and reviews. Additionally, the conference system administrator is responsible of administrating the ConfiChair system for all conferences. Therefore, the *Role* element in the taxonomy is represented in the ConfiChair system scenario. Various roles carry some actions developed by the ConfiChair system. Author upload/download papers, reviewer reviews papers, conference chair referees the process and the conference system administrator administrates all conferences are a few examples. Therefore, the *Action* element in the taxonomy is represented in the ConfiChari system scenario. A set of policies developed within the ConfiChair system. For example, (a) a reviewer doesn't see other reviewers of a paper before writing her own, (b) the Conference System Administrator does not have access to the content of papers or reviews, or the numerical scores give by the reviewers to papers and (c) the conference system administrator does have access to the names of authors and the names of reviewers;

however, he does not have ability to tell if a particular author was reviewed by a particular reviewer. The *Policy* element in the taxonomy is, therefore, represented in the ConfiChair system scenario. The following controls also developed in the ConfiChair system in order to monitor actions: (a) the login procedure implemented relies on each user having an identity *id* and a secret password *pswid;* (b) secrecy of papers, reviews and scores: conference system administrator does not have access to the content of papers or reviews, or the numerical scores given by reviewers to papers; (c) unlinkability of author-reviewer: conference system administrator does not have access to the names of authors and the names of reviewers and (d) however, all sensitive data (e.g. encryption of each review or score) is seen by the conference system administrator only in encrypted form. Therefore, the *Control* element in the taxonomy is also represented in the ConfiChair system scenario. The following evidence such as information produced by uploading/downloading papers, reviewing papers, refereeing the ConfiChair system process and also administrating all conferences through the ConfiChair system is maintained by the ConfiChair system. The *Evidence* element in the taxonomy is therefore represented in the ConfiChari system scenario. Finally, data such as authors, reviewers, conference chairs and conference system administrator information; paper upload/download/submission information; paper review information; and referee process information exist in digital form within the ConfiChair system. Therefore, the *Asset* element in the taxonomy is represented in the ConfiChair system scenario.

The author uploads/downloads (in this case bi-directional) papers and the reviewers reviews (in this case bi-directional) those uploaded papers using the ConfiChair system. Therefore, this creates a trust domain, Author-ConfiChairSystem-Reviewer. According to the ConfiChair system scenario shown in Figure 2, the conference chair referees the process. This creates two trust domains, Author-ConfiChairSystem-ConferenceChair as well as ConferenceChair-ConfiChairSystem-Reviewer.

Finally, the conference system administrator is responsible of administrating the ConfiChair system for all conferences where the conference chair referees the conference process. The conference system administrator does not have access to the names of authors and the names of reviewers and as well as all sensitive data (e.g. encryption of each review or score) is seen by the conference system administrator only in encrypted form. Therefore, this creates a trust domain, ConferenceChair-ConfiChairSystem-ConferenceSystemAdministrator.

## 4. Methodology

The study was mainly focused on a qualitative approach. Qualitative data is the main kind of data used and analyzed by interpretive and critical researchers (Oates, 2006). The study reported in this paper attempted to empirically evaluate the proposed taxonomy for trust domains through the users. This is more likely to be based on the interpretive approach because the study focuses on determining the key elements that should be incorporated into the trust domains taxonomy through the users. Furthermore, it interprets why these elements should be incorporated into the trust domains taxonomy.

### 4.1. Recruiting Participants

An experiment was conducted with 6 participants from the Department of Computer Science at the University of Oxford. All of them were male participants. All of them were academics and researchers in the department. Participants were invited to the Oxford University Computer Science Laboratory. All of them had the experience of being an author (more than five and a half years), reviewer (more than two years) as well as conference chair (more than a year and half). Furthermore, all of them had an experience in using a conference management system for more than five and a half years. Each participant took part in the experiment on a fully voluntarily basis. However, they were offered a cup of coffee in the end of the experiment.

### 4.2. Procedure

We gathered qualitative data from 6 interviews which included observations, field notes and audio recording, and that is, the period of time from the participant arrival to the study and until the participant leaves. All the interviews were conducted in-person by the researcher. First, the nature of the research was explained to each participant individually and they were asked to read and sign the consent form. Then the individual participants were explained the ConfiChair scenario and the difference between the ConfiChair system and exisiting conference management systems such as EasyChair or the Editor's Assistant (EDAS). For example, ConfiChair is a cloud-based conference management protocol developed to address a set of privacy and confidentiality requirements for existing cloud-based conference management systems. Each participant spent approximately one hour with the interview. They were also informed that they could provide any comments and feedback on both the content and format of the study had just been asked to take part.

## 5. Results Analysis

The data analysis of the interviews was conducted in two phrases, which were based on the Norgaard and Hornbaek's study (Nørgaard and Hornbaek, 2006). First, the current study segmented the recordings through the application of keywords to each segment. The purpose of having an interview approach was to empirically evaluate the taxonomy for trust domains through the users using the ConfiChair Scenario. The key words were taken from the elements in the proposed taxonomy as shown in the Figure 1. Donaldson and Fear (Donaldson and Fear, 2011), also stressed in their study that they developed the codeset based on the themes identified in the narratives their subjects provided. Therefore, the audio recordings were mainly segmented into six keywords: *Role, Policy*, *Action, Control, Evidence* and *Asset.* Second, the current study attempted to analysis and tries to form a coherent interpretation of segments that shared keywords. Therefore, the study findings were organized in six areas. Table 1 summarizes these areas and main findings within each of them.

| *Elements of taxonomy* | *Main findings* | *N* | *Example of quotes* |
|---|---|---|---|
| **Role** | Roles such as author, reviewer, conference chair, and conference system administrator are necessary in the ConfiChair system | 6 | "Yes, it is natural thing that you need to have an author, reviewer, conference chair and conference system administrator in any conference management system"<br><br>"I say yes, all roles are important in the conference management system"<br><br>"Yes, we do need roles in any conference system"<br><br>"Well, yes different roles have difference responsibilities"<br><br>"Yes, I believe roles are important, because you want to prevent things like conflict of interests"<br><br>"Yes, it is clear to me that the role is the most important factor, because It may probably have different requirements from the system" |
| **Policy** | Policies developed in the ConfiChair system are important for securely sharing information among others<br>a. A reviewer doesn't see other reviewers of a paper before writing her own<br>b. The conference system administrator does not have access to the content of papers or reviews, or the numerical scores give by the reviewers to papers<br>c. The conference system administrator does have access to the names of authors and the names of reviewers; however, he does not have ability to tell if a particular author was reviewed by a particular reviewer | 6 | "I think all policies developed are important in the system"<br><br>"Yes, it (reviewer sees the other reviews before writing her own) makes biased reviewers"<br><br>"Conference system administrator is the one who provide the overall service and they shouldn't have the details of authors, reviewers and reviewed papers like that"<br><br>"There is a potential that the conference system administrator might change course, modify things, delete things… and you don't want that"<br><br>"Of course, necessary policies should be developed and maintained in order to ensure privacy, security, confidentiality and also integrity of the system"<br><br>"I don't think that the conference system administrator should have access to any of them (author, reviewer, and conference chair) at all"<br><br>"The issue comes down to if the work is currently unpublished, currently in review, then it's about confidentiality, it's about privacy" |
| | | | "It is a desirable thing to have developed policies in place" |
| | | | "No nothing, the conference system administrator should know nothing" |
| | | | "It is a policy that would be useful to establish an unbiased opinion" |
| | | | "Yes, this is something important from privacy perspective, because if I'm an author, I only like people access to my paper who does actually review my paper" |
| **Action** | Actions developed in the ConfiChair system are necessary for securely sharing information among others<br><br>a. Author upload/download papers<br>b. Reviewer reviews papers<br>c. Conference chair referees the process<br>d. Conference System Administrator administrates all conferences | 6 | "Of course, those actions are important in the ConfiChair system"<br><br>"Encryption is important to consider in these actions too"<br><br>"In theory, no you would expect the conference system administrator just to provide a service. However, unless if you encrypt everything on the system they would be able to see"<br><br>"I suppose, when you're uploading/downloading these information (I mean papers, reviews like that) you should know that you're in a secured connection like https. It's about privacy and confidentiality as well"<br><br>"First of all, any sensitive data should be encrypted in all of these actions"<br><br>"All downloading/uploading should be done through https:// in order to protect academics accessing from doggy websites"<br><br>"You should definitely have those actions in place, so you can cater for them. Absolutely, you should be able to cater for the worse case scenarios" |
| **Control** | Controls developed in the ConfiChair system are important for securely sharing information among others<br><br>a. The login procedure implemented relies on each user having an identity id and a secret password pswid<br>b. Secrecy of papers, | 6 | "It's pretty useful and of course highly important to have username and secret password"<br><br>"I imagine that authors, reviewers, and conference chair are all need an unique username and password (of course, for being able to memorise) to login to the system"<br><br>"From the usability prospective I know that you can have the same username and password access to different roles (such as login as author, reviewer or conference chair), but with |
| | reviews and scores: Conference System Administrator does not have access to the content of papers or reviews, or the numerical scores given by reviewers to papers<br><br>c. Unlinkability of author-reviewer: Conference System Administrator does not have access to the names of authors and the names of reviewers<br><br>d. However, all sensitive data (e.g. encryption of each review or score) is seen by the Conference System Administrator only in encrypted form | | "different levels of authentication mechanism"<br><br>"That's what I said before, the conference system administrator should not have access to papers, reviews and score. He just only provides the overall service"<br><br>"To be honest, there is no reason to provide the conference system administrator who the author's name, reviewer's name and etc., because those are quite important services"<br><br>"If the conference system administrator can have access to the database, s/he would be able to pull out the necessary stuff"<br><br>"I think it's better to encrypt all sensitive data on papers, reviewers, reviews, the link between authors and reviewers, stuff like that"<br><br>"I think the only link you should have between the conference system administrator and conference chair, not conference system administrator with others"<br><br>"Yes, papers, reviews and scores should also be kept secret from conference system administrator. Hypothetically, another university can forge some ideas from you"<br><br>"Conference system administrator only really need to see the high level information about the conference such as when is it taking place, who is the main contact, and some technical concerns of the system"<br><br>"If the encryption is done properly, that sounds fine"<br><br>"It is very important authors should not be able to link to their reviewers"<br><br>"The link between the people should be cut, so they can't see each other information"<br><br>"Sure, that's a good practice to keep all sensitive data encrypted form" |
| Evidence | Evidence produced by the ConfiChair system is necessary to maintain within the system<br><br>a. [Provenance] information produced by uploading/downloading papers | 6 | "Yes, the ones [provenance information] you mentioned are important to track within the system"<br><br>"Well, you have to know when the author uploaded the paper in case they passed the deadline and so on"<br><br>"Well, you're talking about the time stamp behind every action which is very important to maintain within the system, e.g. when the |
| | b. [Provenance] information produced by reviewing papers<br><br>c. [Provenance] information produced by refereeing the ConfiChair system process<br><br>d. [Provenance] information produced by administrating all conferences through the ConfiChair system<br><br>For example, provenance information – who, when (date and time), what, and where. | | author uploaded his paper"<br><br>"I think it is important and useful to maintain meta data within the system in case if something goes wrong"<br><br>"Yes, it is important to maintain within the system, because if you're dealing with legal issues, like suddenly someone takes a legal action against the administrator of the system for stealing his/her idea"<br><br>"Yes, you may need this (maintaining the evidence produced by the ConfiChair system) for digital forensic"<br><br>"It actually depends on the role" |
| **Asset** | Assets are important to maintain within the ConfiChair system<br><br>a. Data exist in digital form. For example:<br>  i. Authors, reviewers, conference chairs, and conference system administrator information<br>  ii. Paper upload/ download/ submission information<br>  iii. Paper review information<br>  iv. Referee process information | 6 | "It's important to maintain digital back-up within the system during the conference life time. However, I don't think that you will have to maintain digital back-up after finishing the conference"<br><br>"In terms of maintaining data, it's important to maintain the records of who the authors already submitted, who the reviewed panels are, who reviewed each paper, actually papers as well. But not all information necessarily be maintained in digital form"<br><br>"You certainly need to keep it for the duration of the conference. After that you may probably need to destroy"<br><br>"It is nice to keep the data in a back-up form and I see the purpose of it here"<br><br>"During the conference, yes I will say so. If anyone has any issues, you can provide the evidence here"<br><br>"From security perspective, three big letters are CIA, in terms of Integrity, yes you do need to maintain or keep tract of these assets. |

**Table 1: Overview of results - N refers to the number of participants (out of 6 interview sessions in total)**

## 6. Trust Domains Taxonomy

The current study empirically investigated what key elements that should be incorporated into the trust domains taxonomy for securely sharing information among others. In addition, it interprets why these elements should be incorporated into the trust mains taxonomy. The elements of the trust domains taxonomy introduced by Arachchilage, et al. (Arachchilage *et al.* 2013), were used to empirically investigate through the ConfiChair system scenario. Therefore, a qualitative study was conducted to assess the taxonomy. The study employed 6 participants (as a pilot study) with each participant participating for an approximately one-hour session.

All participants talked about their opinions about the elements of the taxonomy through the ConfiChair scenario. All of them believed that the protocol underlying ConfiChair, a novel cloud-based conference management system, offers strong privacy and confidentiality guarantees. Furthermore, they talked about their opinions of the elements of the taxonomy as shown in Figure 1 through the ConfiChair system scenario. The taxonomy elements are *Role, Policy*, *Action, Control, Evidence* and *Asset*. The study revealed that *Role, Policy*, *Action, Control, Evidence* and *Asset* elements should be incorporated into the taxonomy. Therefore, the current study findings provided evidence of addressing the above elements in the trust domains taxonomy for securely sharing information among others.

## 7. Discussion

The current study empirically investigated what key elements that should be incorporated into the trust domains taxonomy for securely sharing information among others. Furthermore, it determines why these elements should be incorporated into the trust domains taxonomy.

All participants talked about their opinions of the elements of the taxonomy through the ConfiChair scenario. All of them were convinced in our pilot study that the taxonomy (shown in Figure 1) is somewhat effective in securely sharing information among others. Their common argument was that the conference system administrator should not have access to the names of authors and the names of reviewers including papers and reviews. Additionally, they argued that all sensitive data (e.g. encryption of each review or score) is seen by the conference system administrator should only be in encrypted form. One participant responded, *"No nothing, the conference system administrator should know nothing"*. Therefore, the current study conveys a simple, yet a powerful message that the proposed ConfiChair system offers strong privacy and confidentiality guarantees (Arapinis *et al.* 2012).

All participants were believed that the *Role* element is important in the taxonomy. *Roles* are used to specify the level of participation in the ConfiChair system. One participant responded, "*Yes, it is natural thing that you need to have an author, reviewer, conference chair and conference system administrator in any conference management system*". It is true that the ConfiChair system provides the facility to different system requirements for its intended users such as an author, reviewer, conference chair, and the conference system administrator. In terms of the confidentiality and privacy perspective, it is, therefore, worth considering the conflict of interests or requirements of different user roles. One participant stressed that,

*"Yes, I believe roles are important, because you want to prevent things like conflict of interests"*. It can therefore be argued that the *Role* element should be incorporated into the trust domains taxonomy for securely sharing information among others.

A *Role* establishes one or more *Policies* within the trust domain. *Policies* are a means of specifying the behavior of entities within the ConfiChair system. All participants believed that given *Policies* developed in the ConfiChair system are important for securely sharing information among others (in this case for strong privacy and confidentiality gurantees). For example, a reviewer doesn't see other reviewers of a paper before writing her own. All participants believed that this policy prevents the ConfiChair system from being biased reviewers. One participant commented on the above policy, *"It is a policy that would be useful to establish an unbiased opinion"*. Furthermore, the proposed ConfiChair system has another policy in place called the conference system administrator does not have access to the content of papers or reviews, or the numerical scores given by the reviewers to papers. This is imperative to offer strong privacy and confidentiality guarantees in the ConfiChair system. One participant said. *"Of course, necessary policies should be developed and maintained in order to ensure privacy, security, confidentiality and also integrity of the system"*. Another participant also stressed, *"The issue comes down to if the work is currently unpublished or in-review, then it's about confidentiality, it's about privacy"*. These statements describe how much developed policies are important in the ConfiChair system. Therefore, the *Policy* element is significantly important to incorporate into the trust domains taxonomy for securely sharing information among others.

The trust domains taxonomy describes *Actions* are performed by a given *Role* or by an agent that represents a particular *Role*. *Actions* are a series of functionalities performed by author, reviewer, conference chair, or conference system administrator in the ConfiChair system. Author uploads papers or download reviews, reviewer reviews papers, conference chair referees the process and the conference system administrator administrates all conferences are a few examples for *Actions* in the ConfiChair system. All participants believed that these actions are important in the ConfiChair system. One participant stated, *"I suppose, when you're uploading/downloading these information (I mean papers, reviews like that) you should know that you're in a secured connection like https. It's about privacy and confidentiality as well"*. It is important that the end-user should perceive the trust of his or her sensitive data such as username, password, and paper information from the proposed ConfiChair system. Another participant said, *"First of all, any sensitive data should be encrypted in all of these actions"*. These statements describe how much developed *Actions* are important in the ConfiChair system. Therefore, the *Action* element is significantly important to incorporate into the trust domains taxonomy for securely sharing information among others.

The trust domains taxonomy shown in Figure 1 describes *Controls* are a set of mechanisms, processes or procedures that enforce the *Policies* within a trust domain. These controls could be accomplished through social, e.g. penalties, or technical means e.g. trusted computing. Controls monitor activities that occur within a trust domain and produce evidence, described below, that can be used to determine the properties of a trust domain or its constituents. All participants believed that a set of controls developed in the ConfiChair system are important for securely sharing information among others. One participant commented, *"It's pretty useful and of course highly important to have username and secret password"*. Another participant backed up saying, *"From the usability prospective I know that you can have the same username and password access to different roles (such as login as author, reviewer or conference chair) but with different levels of authentication mechanism"*. It is worth understanding how these controls accomplish through technical means are important for securely sharing information among others. Another participant stated, *"I think it's better to encrypt all sensitive data on papers, reviewers, reviews, the link between authors and reviewers, stuff like that"*. This provides further evidence for these controls accomplish through technical means are necessary for securely sharing information among others. It can therefore be argued that the *Control* element is significantly important to incorporate into the trust domains taxonomy for securely sharing information among others.

The trust domains taxonomy shown in Figure 1 describes *Evidence* is data that is produced by the controls within a trust domain to indicate the kinds of activities that have occurred in a trust domain. These activities are captured by monitoring the actions that are performed by or on behalf of roles that exists within a trust domain. For example, such evidence can be provenance - records of how data came to be. All participants believed that *Evidence* produced by the ConfiChair system is necessary to maintain within the system. Examples of such evidence include; provenance information produced by uploading/downloading papers, reviewing papers, refereeing the ConfiChair system process and administrating all conferences through the ConfiChair system. One participant responded, *"Well, you have to know when the author uploaded the paper in case they passed the deadline and so on"*. On the other hand, it is very useful to maintain some evidence for digital forensic purposes. One participant said, *"Yes, it is important to maintain [provenance information] within the system, because if you're dealing with legal issues, like suddenly someone takes a legal action against the administrator of the system for stealing his/her idea"*. These statements describe how much worth of maintaining *Evidence* the ConfiChair system. Therefore, the *Evidence* element is significantly important to incorporate into the trust domains taxonomy for securely sharing information among others.

Conceptualization of a trust domain is based on the idea of enabling secure information flow among a set of entities. Such entities may each have a set of devices through which they share the data. Furthermore, these entities may provide access to the information stored on the devices or other media to other members of the domain. For this reason, Arachchilage, et al. (Arachchilage *et al.* 2013), define the concept of an *Asset* as being a fundamental element of a trust domain. An asset is something of value to the owner, but could also be valuable to other entities such as attackers or competitors. One example of an asset is data. All participants agreed that the data exist in digital form should be maintained during the period of the particular conference. They also believed if an issue arose, then you may need this information as evidence. For example, if an author made an inquiry asking to double-check the feedback of the reviewer, in case if he has received someone else's feedback. One participant responded, *"During the conference, yes I will say so. If anyone has any issues, you can provide the evidence here"*. However, all participants believed that it is not necessary to maintain data after the conference with respect to the ConfiChair system. For example, one participant said, *"You certainly need to keep it for the duration of the conference. After that you may probably need to destroy"*. It can therefore be argued that it is important to maintain data within the system during its lifetime. This is because certain data could have detrimental effects when it ends up in wrong hands like hackers. Therefore, the *Asset* element is significantly important to incorporate into the trust domains taxonomy for securely sharing information among others.

## 8. Conclusions and Future work

This research attempted to empirically investigate the proposed taxonomy through the ConfiChair system scenario. The study asks what elements that should be incorporated into the trust domains taxonomy for securely sharing information among others and why those elements are important. Finally, the current study results provided support to define and justify what key elements that should be addressed in the trust domains taxonomy for securely sharing information among others. The study results showed that *Role, Policy*, *Action, Control, Evidence* and *Asset* elements should be incorporated into the taxonomy for securely sharing information among others. Additionally, the study findings revealed that the protocol underlying ConfiChair, a novel cloud-based conference management system, offers strong privacy and confidentiality guarantees.

This study has identified some limitations. First, we interviewed 6 participants as a pilot study from the Department of Computer Science at the University of Oxford. Future research is needed to confirm our findings using different samples (relatively with a large sample size). Furthermore, it is worthwhile reporting the main study relatively with a large sample size on the quality of the taxonomy, in terms of its completeness, integrity, flexibility, understandability, correctness, simplicity, integration, and implementation as suggested by Moody and Shanks (Moody and Shanks, 2003). Second, for the purpose of empirical testing, we selected ConfiChair system scenario as the possible scenario. Therefore, research can be conducted with different scenarios to examine whether or not the findings of this study will change.

## 9. References

Arachchilage, N. A. G., Namiluko, C. and Martin, A. (2013). "A Taxonomy for Securely Sharing Information Among Others in a Trust Domain," *8th International Conference for Internet Technology and Secured Transactions (ICITST)*, London, IEEE, vol., no., pp.296,304.

Arapinis, M., Bursuc, S. and Ryan, M. (2012). "Privacy supporting cloud computing: ConfiChair, a case study," *Principles of Security and Trust*, Springer, pp. 89–108.

Baptista, A., Howe, B., Freire, J., Maier, D. and Silva, C. T. (2008). "Scientific exploration in the era of ocean observatories," *Comput. Sci. Eng.*, Vol. 10, No. 3, pp. 53–58.

Bizer, C. and Oldakowski, R. (2004). "Using context-and content-based trust policies on the semantic web", *Proceedings of the 13th international World Wide Web conference on Alternate track papers & posters*, pp. 228–229.

Chapman, A., Blaustein, B. and Elsaesser, C. (2010). "Provenance-based belief," In *Proceedings of the 2nd conference on Theory and practice of provenance* (TAPP'10). USENIX Association, Berkeley, CA, USA, pp. 11-11.

Donaldson, D. R. and Fear, K. (2011). "Provenance, end-user trust and reuse: An empirical investigation," in *TaPP'11: Proceedings of the third USENIX Workshop on the Theory and Practice of Provenance*, Heraklio, Crete, Greece.

Gil, Y. and Artz, D. (2007). "Towards content trust of web resources," *Web Semant. Sci. Serv. Agents World Wide Web*, Vol. 5, No. 4, pp. 227–239.

Hartig, O. and Zhao, J. (2009). "Using Web Data Provenance for Quality Assessment," *SWPM*, Vol. 526.

Kelton, K., Fleischmann, K. R. and Wallace, W. A. (2008). "Trust in digital information," *J. Am. Soc. Inf. Sci. Technol.*, Vol. 59, No. 3, pp. 363–374.

Kirkham, T. Winfield, S. Ravet, S. and Kellomaki, S. (2013). "The Personal Data Store Approach to Personal Data Security," *IEEE Secur. Priv.*, Vol. 11, No. 5, pp. 12–19.

Lauriault, T. P., Craig, B. L., Taylor, D. R. and Pulsifer, P. L. (2008). "Today's data are part of tomorrow's research: Archival issues in the sciences," *Archivaria*, Vol. 64.

Moody, D. L. and Shanks, G.G. (2003). "Improving the quality of data models: empirical validation of a quality management framework", Information Systems Journal, vol. 28, pp. 619-650.

Missier, P., Belhajjame, K., Zhao, J., Roos, M. and Goble, C. (2008). "Data lineage model for Taverna workflows with lightweight annotation requirements," *Provenance and Annotation of Data and Processes*, Springer, pp. 17–30.

Nørgaard, M. and Hornbæk, K. (2006). "What Do Usability Evaluators Do in Practice?: An Explorative Study of Think-aloud Testing," *Proceedings of the 6th Conference on Designing Interactive Systems*, New York, NY, USA, 2006, pp. 209–218.

Sexton, A., Yeo, G., Turner, C. and Hockey, S. (2004). "User feedback: testing the LEADERS demonstrator application," *J. Soc. Arch.*, Vol l. 25, No. 2, pp. 189–208.

Wang, Y. D. and Emurian, H. H. (2005). "An overview of online trust: Concepts, elements, and implications," *Comput. Hum. Behav.*, Vol. 21, No. 1, pp. 105–125.

Weng, C., Gennari, J. H. and Fridsma, D. B. (2007). "User-centered semantic harmonization: A case study," *J. Biomed. Inform.*, Vol. 40, No. 3, pp. 353–364.